\documentclass[conference]{IEEEtran}
\IEEEoverridecommandlockouts

\usepackage{cite}
\usepackage{amsmath,amssymb,amsfonts}
\usepackage{algorithmic}
\usepackage{graphicx}
\usepackage{textcomp}
\usepackage[version=4]{mhchem}
\usepackage{booktabs}
\usepackage{multirow}
\usepackage{siunitx}
\usepackage{mhchem}
\usepackage{amssymb}
\usepackage{multirow}
\usepackage{graphicx}
\usepackage{eurosym}
\def\BibTeX{{\rm B\kern-.05em{\sc i\kern-.025em b}\kern-.08em
    T\kern-.1667em\lower.7ex\hbox{E}\kern-.125emX}}
\usepackage{url}
\usepackage{subcaption} 

\usepackage{tikz}
\usetikzlibrary{calc,arrows,patterns,intersections}
\usepackage{pgfplots}
\usepackage{bbm}
\usepackage{bm}
\usepackage{adjustbox}
\usepackage{xcolor}
\usepgfplotslibrary{fillbetween}

\usepackage{alphalph}
\usepackage{etoolbox}

\AtBeginDocument{%
  \AtBeginEnvironment{subequations}{%
    \let\alph\alphalphval%
  }
}

\definecolor{RectangleColor}{RGB}{197,210,180}

\begin{document}

\title{Energy-Intensive Industries Providing Ancillary Services: A  Real Case of Zinc Galvanizing Process}

\author{Peter A.V. Gade\textsuperscript{*}\textsuperscript{\textdagger}, Trygve Skjøtskift\textsuperscript{\textdagger}, Henrik W. Bindner\textsuperscript{*}, Jalal Kazempour\textsuperscript{*} \\
    \textsuperscript{*}Department of Wind and Energy Systems, Technical University of Denmark, Kgs. Lyngby, Denmark \\
    \textsuperscript{\textdagger}IBM Client Innovation Center, Copenhagen, Denmark
    \thanks{
        Email addresses: pega@dtu.dk (P.A.V. Gade), Trygve.Skjotskift@ibm.com (T. Skjøtskift), hwbi@dtu.dk (H.W. Bindner), jalal@dtu.dk (J. Kazempour). The authors would like to acknowledge the financial support from Innovation Fund Denmark under grant number 0153-00205B. The authors would also like to thank DOT Nordic for giving access to their factory and providing the data used in this paper.}

    \vspace{-3mm}
}

\maketitle

\IEEEaftertitletext{\vspace{-0.8\baselineskip}}
\maketitle

\begin{abstract}
    Energy-intensive industries can adapt to help balance the power grid. 
    By using a real-world case study of a zinc galvanizing process in Denmark, we show how a modest investment in power control of the furnace enables the provision of various ancillary services. We consider two types of services, namely frequency containment reserve (FCR) and manual frequency restoration reserve (mFRR), and numerically conclude that the monetary value of both services is significant, such that the pay-back time of investment is potentially within a year. The FCR service provision is more preferable as its impact on the temperature of the zinc is negligible.
\end{abstract}

\vspace{1mm}
\begin{IEEEkeywords}
    Demand-side flexibility, thermostatically controlled loads, ancillary services
\end{IEEEkeywords}

\section{Introduction}
\vspace{-1mm}
Power prices in Europe increased massively in 2022. Meanwhile, the increasing share of renewable energy sources with stochastic production stresses the grid to an unprecedented level. In 2022, the Danish transmission system operator (TSO), Energinet, spent 2.7 billion DKK for the procurement of various ancillary services, showing an increase of 1.3 billion DKK from 2021 \cite{energinetOmkostninger}. Energinet expects this cost to be increased further as the integration of additional stochastic renewables requires more balancing resources.

It is more appealing to exploit existing infrastructure to supply power flexibility and  meet the increasing demand for balancing resources. In particular, energy-intensive consumers can become flexible to some degree, i.e., their power consumption can vary within some thresholds for a certain amount of time as facilitated by a flexibility aggregator. IBM has created a platform, so called \textit{Flex Platform}, that harnesses demand-side flexibility to be offered into various ancillary service markets. One may hypothesize a modest investment making industrial plants  to some extent flexible is financially advantageous for their owners. This paper explores it with a particular focus on industries with \textit{single-state} processes. 

Single-state processes are characterized by being solely sequential, i.e., the next step occurs after the previous step is done. They are repetitive as well, relatively simple and uncomplicated. Their state is often a temperature level. They exist everywhere: iron and steel foundries, cooling houses, and a zinc galvanizing operation as it is the case study of this paper. See \cite{paulus2011potential} for an extensive list of energy-intensive industry processes. Such processes are prone to become flexible with respect to their power consumption as it requires little to no extra effort or adjustment. Furthermore, it could be more desirable to harness the flexibility from one big asset as supposed to several hundreds smaller assets\footnote{This might not be the case for flexibility required for grid congestion management, while for frequency services needed for balancing the system, the location does not necessarily matter. Our focus in this paper is on flexibility resources for frequency services.}.


In this work, an industry process is investigated as exemplified by a zinc galvanizing furnace using real data from their process. We investigate how it is feasible for the zinc furnace to become a flexible energy consumer and provide two common types of ancillary services, namely frequency containment reserve (FCR) and manual frequency restoration reserve (mFRR). We conduct a techno-economic analysis while investigating the temperature impact of delivering flexibility.  The learnings from this case study can be readily generalized to other single-state processes.

\vspace{0mm}
\subsection{Research questions and context}
\vspace{-1mm}
We address four research questions: (\textit{i}) are there incentives for an energy-intensive industry process to provide power flexibility to the grid? (\textit{ii}) how can it adapt its industry process to provide power flexibility? (\textit{iii}) are ancillary services such as FCR and mFRR suited for such a process? and eventually (\textit{iv})  what is the monetary impact of temperature thresholds?



New investments into equipment and improvements in the industry process can be tailored to facilitate power flexibility. For example, instead of relying on mechanical relay switches for power lines, frequency transformers or thyristors can be used with a modest investment cost. The pay-back time for such an investment should ideally be within a few years, and other benefits could potentially be exploited as well, e.g., better feed-forward planning.
%
%
%
Power consumption to heat up zinc can be controlled as long as the temperature level of the zinc is within some pre-specified thresholds. This determines the degree of freedom in the operation. For the zinc furnace, reaching to the lower temperature threshold leads to the solidification of molten zinc which can have a detrimental impact as it can crack the furnace wall.
The ability to deviate from the baseline power consumption can be monetized by participating in FCR and mFRR markets. 



\vspace{1mm}
\subsection{Literature review, our contributions, and outline}
\vspace{-1mm}
Many studies have investigated demand-side flexibility. While most have focused on flexibility from residential households in an aggregated portfolio, some such as \cite{paulus2011potential} have looked at energy-intensive industry processes. 
According to the best of our knowledge, none specifically looks at single-state industrial processes in a real-life setting while addressing their incentives to deliver power flexibility.


%
It is argued in \cite{mallapragada2023decarbonization} that energy-intensive chemical processes can benefit from investments into electrification. The payback time of the investments can be reduced significantly by exploiting the capabilities of power electronics to deliver ancillary services to the grid. Likewise, it is shown in \cite{samani2022flexible} how a chemical process can deliver FCR to the grid while still maintaining the same operational quality.
In \cite{junker2020stochastic}, a water tower is characterized using stochastic differential equations for optimally planning how water should be pumped in relation to spot prices. Although not specifically for ancillary services, it serves as an illustrative example how operational control can benefit from knowledge of electricity prices in a fairly simple industry process.


As our main contribution, for a real-life energy-intensive industry process using real data, we investigate if there is any incentive to provide power flexibility or to invest in equipment, enabling the flexibility provision.
Specifically, we investigate such a case using actual (anonymized) data collected from 2022 as well as price and frequency data for 2021-2023. We describe the process in detail and show how it can be adapted to cater for power flexibility by switching from ON/OFF power control to continuous power control.
We finally provide an upper bound of flexibility earnings using a full hindsight optimization of each service and its impact on the temperature. This means we analyze the state of the industry process and how it is impacted by the flexibility provision. We discuss how the allowed degrees of freedom in the temperature deviation impact the monetary value of flexibility provision.
Our case study naturally generalizes to many similar single-state industry processes seen in, e.g., foundries where the state in question is temperature related.


The rest of the paper is organized as follows. 
Section \ref{sec:zinc_furnace_description} provides preliminaries explaining how to characterize the zinc furnace as a flexible load.
Section \ref{opt} proposes two optimization problems for the zinc furnace owner to bid in either FCR or mFRR markets. Section \ref{results} provides numerical results. Finally, Section \ref{conc} concludes the paper.



\vspace{0.5mm}
\section{Preliminaries: \\ Characterizing the zinc furnace}\label{sec:zinc_furnace_description}

\vspace{-1mm}
%
This section describes the zinc galvanizing process and how to characterize the zinc furnace as a thermostatically controlled load (TCL). In particular, we first explain how the temperature of the molten zinc can be characterized using a state-space model. Then, it is discussed how this model can be used for a subsequent simulation of continuous control. 

A straightforward mathematical model of a TCL is presented in  \cite{hao2014aggregate}, where a first-order model with two terms characterizes the TCL. It includes a term that explains temperature losses due to temperature differences and a term that explains temperature gains due to a power source. We expand upon this model, using instead a 4$^{\text{th}}$ order model to characterize a zinc furnace. While a linear model is used for modeling FCR provision in this paper, the model introduced in \cite{gade2023load} is used to model mFRR provision.

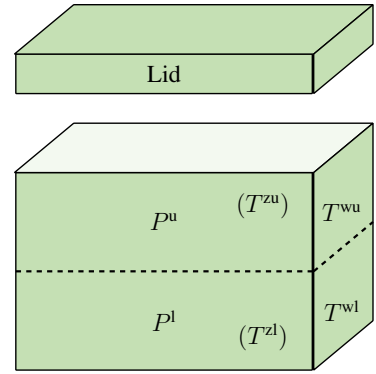
\begin{figure}[t]
    \centering
    \begin{adjustbox}{width=0.55\columnwidth}
        \begin{tikzpicture}
  \definecolor{CUBE}{RGB}{197,224,180};
  \coordinate (CenterPoint) at (0,0);
  \def\width{6.0cm};
  \def\height{4.0cm};
  \def\textborder{1.0cm};
  \def\xslant{1.2cm};
  \def\yslant{1.0cm};
  \def\rounding{0.4pt};
  \node[thick, draw,
    minimum height  = \height,
    minimum width   = \width,
    text width      = {\width-1*\textborder},
    align           = center,
    fill            = CUBE,
    rounded corners = \rounding]
  at (CenterPoint) {}; 

  \draw[name path=HorizontalDashedLine, dashed, line width=0.45mm, black]%
  ($(CenterPoint) + (-\width/2., 0)$) -- ($(CenterPoint) + (\width/2., 0)$);

  \draw [rounded corners = \rounding, thick, fill=CUBE!20] %
  ($(CenterPoint) + (-\width/2. - 2*\rounding, \height/2.)$) -- %
  ($(CenterPoint) + (-\width/2. + \xslant - 2*\rounding, \height/2. + \yslant)$) -- %
  ($(CenterPoint) + (\width/2. + \xslant + 2*\rounding, \height/2. + \yslant)$) -- %
  ($(CenterPoint) + (\width/2. + 2*\rounding, \height/2.)$) -- %
  cycle;
  \draw [rounded corners = \rounding, thick, fill=CUBE] %
  ($(CenterPoint) + (\width/2. + \xslant + 2*\rounding, \height/2. + \yslant)$) -- %
  ($(CenterPoint) + (\width/2. + 2*\rounding, \height/2.)$) -- %
  ($(CenterPoint) + (\width/2. + 2*\rounding, -\height/2.)$) -- %
  ($(CenterPoint) + (\width/2. + \xslant + 2*\rounding, -\height/2. + \yslant)$) -- %
  cycle;

  \coordinate (CenterPointLid) at (0,4);
  \node[thick, draw,
    minimum height  = \height*0.2,
    minimum width   = \width,
    text width      = {\width-1*\textborder},
    align           = center,
    fill            = CUBE,
    rounded corners = \rounding]
  at (CenterPointLid){\Large Lid};

  \draw [rounded corners = \rounding, thick, fill=CUBE] %
  ($(CenterPointLid) + (-\width/2. - 2*\rounding, 0.2*\height/2.)$) -- %
  ($(CenterPointLid) + (-\width/2. + \xslant - 2*\rounding, 0.2*\height/2. + \yslant)$) -- %
  ($(CenterPointLid) + (\width/2. + \xslant + 2*\rounding, 0.2*\height/2. + \yslant)$) -- %
  ($(CenterPointLid) + (\width/2. + 2*\rounding, 0.2*\height/2.)$) -- %
  cycle;

  \draw [rounded corners = \rounding, thick, fill=CUBE] %
  ($(CenterPointLid) + (\width/2. + \xslant + 2*\rounding, 0.2*\height/2. + \yslant)$) -- %
  ($(CenterPointLid) + (\width/2. + 2*\rounding, 0.2*\height/2.)$) -- %
  ($(CenterPointLid) + (\width/2. + 2*\rounding, -0.2*\height/2.)$) -- %
  ($(CenterPointLid) + (\width/2. + \xslant + 2*\rounding, -0.2*\height/2. + \yslant)$) -- %
  cycle;

  \draw[name path=HorizontalDashedLineSide, dashed, line width=0.45mm, black]%
  ($(CenterPoint) + (\width/2. + 2*\rounding, 0)$) -- ($(CenterPoint) + (\width/2. + \xslant + 2*\rounding, + \yslant)$);

  \node at ($(CenterPoint) + (\width/2. + \xslant/2., \yslant*1.2)$) %
  {\Large $T^{\text{wu}}$};
  \node at ($(CenterPoint) + (\width/2. + \xslant/2., -\yslant*0.8)$) %
  {\Large $T^{\text{wl}}$};

  \node at ($(CenterPoint) + (\width/3., \height/3)$) %
  {\Large $(T^{\text{zu}})$};
  \node at ($(CenterPoint) + (\width/3., -\height/3)$) %
  {\Large $(T^{\text{zl}})$};

  \node at ($(CenterPoint) + (0, \height/4)$) %
  {\Large $P^{\text{u}}$};
  \node at ($(CenterPoint) + (0, -\height/4)$) %
  {\Large $P^{\text{l}}$};

\end{tikzpicture}
    \end{adjustbox}
    \caption{\small{Schematic of the zinc furnace. 
            \vspace{-4mm}
    }}
    \label{fig:furnace_schematic_tikz}
\end{figure}

The industry process exemplified in this paper is a galvanization process, in which the authors have kindly been allowed to visit and learn about by the owner, DOT Nordic. Steel elements are galvanized by lowering them into a molten zinc furnace at a high temperature setpoint. As illustrated in Fig. \ref{fig:furnace_schematic_tikz}, the furnace is heated up by resistive elements placed on the sides of the furnace (in an inner cavity). The upper and lower zones of the furnace are controlled separately with $P^{\text{u}}$ and $P^{\text{l}}$, representing the power supplied to the upper and lower zones, respectively. Temperature sensors are placed in each zone at the end of the furnace on the wall as denoted by $T^{\text{wu}}$ and $T^{\text{wl}}$, hence the actual zinc temperatures, $(T^{\text{zu}})$ and $(T^{\text{zl}})$, are latent, unobserved states. The lid of the furnace is removed when lowering steel into the furnace. We denote QU1 and QU2 as the contactors for the upper zone, whereas QL3 and QL4 are the contactors for the lower zone.

\begin{figure}[t]
    \centering
    \includegraphics[width=0.90\columnwidth]{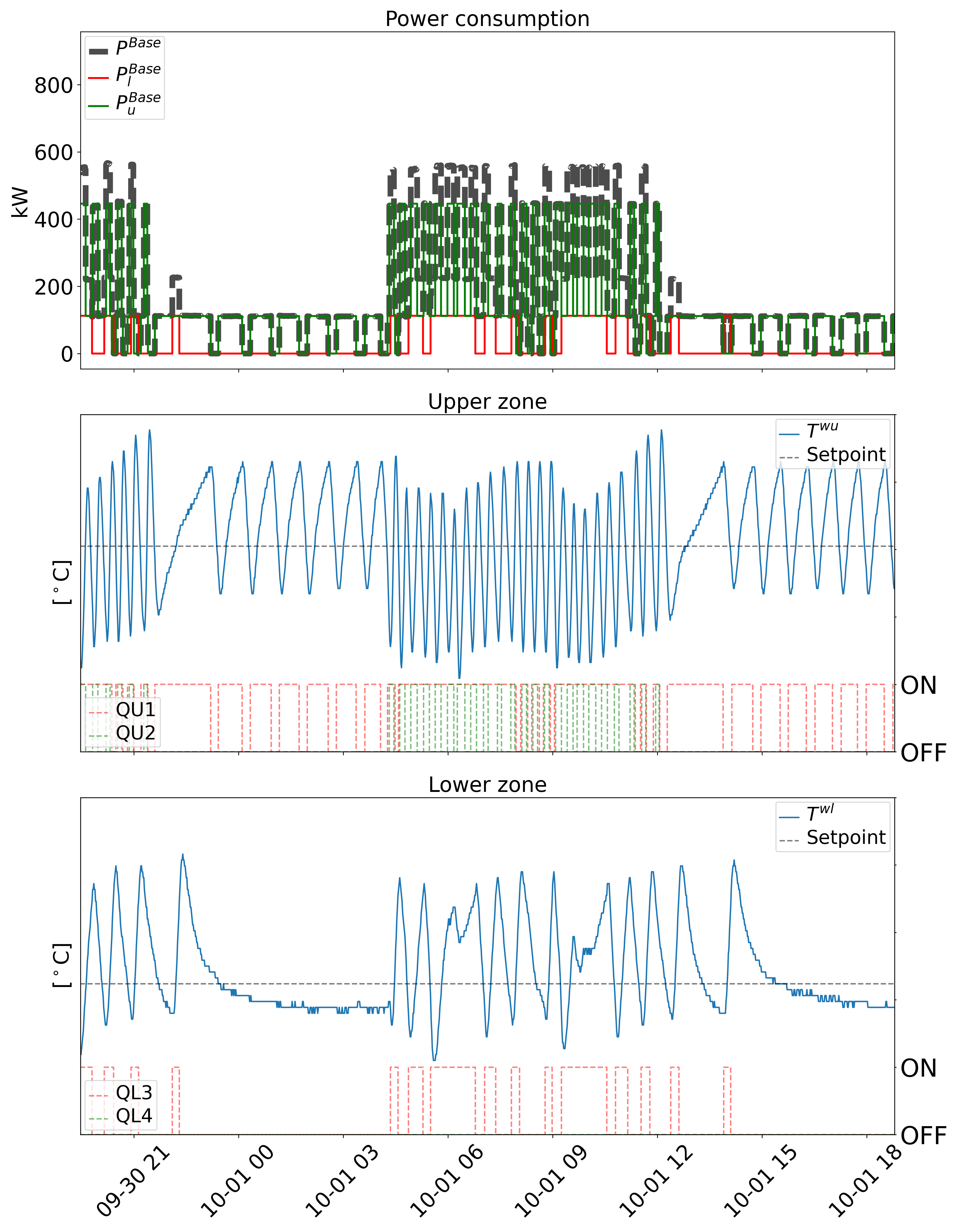}
    \caption{\small{\textbf{Top}: Power consumption for lower and upper zones and the total consumption. \textbf{Middle}: Temperature and contactor switches in the upper zone. \textbf{Bottom}: The same for the lower zone.
                \vspace{-4mm}}}
    \label{fig:data_visualization}
\end{figure}

Fig. \ref{fig:data_visualization} shows the one-minute resolution data. The scale of the temperature has been anonymized. The upper plot shows the power consumption of the lower and upper zones, respectively, and the total consumption as well. It is clearly observed how two regimes are immediately identified: One when the lid is on the furnace and the power consumption is quite low, and another one when the lid is off corresponding to a high power consumption due to direct temperature losses to the ambient. In both middle and bottom plots of Fig. \ref{fig:data_visualization}, the temperature dynamics behave differently in either regime: When the lid is on, the temperature varies slowly while it varies rapidly when the lid is off. Furthermore, the temperature dynamics are generally slower in the lower zone of the furnace (bottom plot), and its temperature setpoint is also a bit lower. The ON/OFF control of the power consumption to the two zones are visualized by the state of four contactors -- two for each zone -- as depicted in the middle and bottom plots. The control logic is simple for a given zone: one contactor is switched ON (QU1 $=1$ or QL3 $=1$) when the temperature goes below a pre-specified threshold, and the other one is turned ON (QU2 $=1$ or QL4 $=1$) as well if the temperature declines further below another pre-specified threshold. The same logic applies when the temperature rises above two pre-specified upper thresholds. This logic is statically programmed for both zones independently.

\vspace{0.2mm}
\subsection{Thermal modeling of the zinc furnace}
\vspace{-1mm}
To model the temperature dynamics in both zones, a fourth-order state-space model is developed as
%
\begingroup
\allowdisplaybreaks
\begin{subequations}\label{eq:StateSpaceModel}
    \begin{align}
        T^{\text{zu}}_{t+1} & = T^{\text{zu}}_{t} +   \frac{dt}{C^{\text{zu}}}\Bigl( \frac{1}{R^{\text{zuzl}}} (T^{\text{zl}}_{t} - T^{\text{zu}}_{t}) + \frac{1}{R^{\text{wz}}} (T^{\text{wu}}_{t} - T^{\text{zu}}_{t}) \Bigr) \label{eq1:StateSpaceModel} \\
        T^{\text{zl}}_{t+1} & = T^{\text{zl}}_{t} +   \frac{dt}{C^{\text{zl}}}\Bigl( \frac{1}{R^{\text{zuzl}}} (T^{\text{zu}}_{t} - T^{\text{zl}}_{t})  + \frac{1}{R^{\text{wz}}} (T^{\text{wl}}_{t} - T^{\text{zl}}_{t}) \Bigr) \label{eq2:StateSpaceModel} \\
        T^{\text{wu}}_{t+1} & = T^{\text{wu}}_{t} +   \frac{dt}{C^{\text{wu}}}\Bigl( (1-\mathbbm{1}^{\text{lid}}) \frac{1}{R^{\text{wua},\text{off}}} (T^{\text{a}} - T^{\text{wu}}_{t}) \notag + \\ & \mspace{5mu} \mathbbm{1}^{\text{lid}} \frac{1}{R^{\text{wua},\text{on}}} (T^{\text{a}} - T^{\text{wu}}_{t}) + \frac{1}{R^{\text{ww}}} (T^{\text{wl}}_{t} - T^{\text{wu}}_{t}) \notag \\ & \mspace{5mu} + \frac{1}{R^{\text{wz}}} (T^{\text{zu}}_{t} - T^{\text{wu}}_{t}) + p^{\text{u}}_{t} \Bigr) \label{eq3:StateSpaceModel} \\
        T^{\text{wl}}_{t+1} & = T^{\text{wl}}_{t} +   \frac{dt}{C^{\text{wl}}}\Bigl( \frac{1}{R^{\text{wla}}} (T^{\text{a}} - T^{\text{wl}}_{t}) \notag                                           \\ & \mspace{5mu} + \frac{1}{R^{\text{ww}}} (T^{\text{wu}}_{t} - T^{\text{wl}}_{t}) + \frac{1}{R^{\text{wz}}} (T^{\text{zl}}_{t} - T^{\text{wl}}_{t}) + p^{\text{l}}_{t} \Bigr), \label{eq4:StateSpaceModel}
    \end{align}
\end{subequations}
\endgroup
where the heat capacities of the wall and the zinc for both upper and lower zones are denoted as $C^{\text{wl}}$, $C^{\text{wu}}$, $C^{\text{zl}}$, and $C^{\text{zu}}$, respectively. Furthermore, there are two different resistance coefficients, $R^{\text{wua},\text{off}}$ and $R^{\text{wua},\text{on}}$, depending on whether the lid is on ($ \mathbbm{1}^{\text{lid}} = 1$) or off ($ \mathbbm{1}^{\text{lid}} = 0$). The index $t$ represents a minute and $dt = \frac{1}{60}$ is the time step. The remaining resistant coefficients are related to heat transfer within the furnace wall, $R^{\text{ww}}$, heat transfer from the wall to the zinc, $R^{\text{wz}}$, heat transfer from the wall to the ambient, $R^{\text{wa}}$, and heat transfer within the zinc, $R^{\text{zuzl}}$. Equations (\ref{eq:StateSpaceModel})  represent the (latent) temperature of the zinc in the upper and lower zones. For both, there is a heat exchange between the furnace walls and the other zinc zone. In (\ref{eq3:StateSpaceModel}), the temperature dynamics is modeled for the upper part of the wall of the zinc furnace. It depends on the heat loss to the ambient temperature ($T^{\text{a}}$), heat exchange with the lower wall ($T^{\text{wl}}$), with the zinc in the upper zone ($T^{\text{zu}}$), and the heat added from the resistive elements in the upper zone  ($p^{\text{u}}$).   The procedure of parameter estimation in (\ref{eq:StateSpaceModel}) is explained in Appendix A.

\vspace{1mm}
\subsection{Moving from ON/OFF to steady-state control}
\vspace{-1mm}
As observed in Fig. \ref{fig:data_visualization}, the power consumption is rather unpredictable and varying with the ON/OFF control mechanism. It is difficult to predict the operational baseline power consumption of the zinc furnace. Therefore, if the furnace should benefit from providing flexibility to the power grid through ancillary services, it is necessary to change the control structure to a more granular one. In theory, the furnace can still participate in the current market structure of mFRR with 1-hour blocks if it is part of a bigger portfolio that can compensate for its ON/OFF behavior.

Using (\ref{eq:StateSpaceModel}), we  estimate the steady-state power consumption for both regimes, i.e., when the lid is on and off, as
%
%
\begin{equation}\label{eq1:steady-state-power}
    p^{\text{l},\text{Base}} = \frac{T^{\text{l},\text{sp}} - T^{\text{a}}}{R^{\text{wla}}} - \frac{T^{\text{u},\text{sp}}-T^{\text{l},\text{sp}}}{R^{\text{ww}}}
\end{equation}

\begin{equation}\label{eq2:steady-state-power}
    p^{\text{u},\text{Base}}   =
    \begin{cases}
        \frac{T^{\text{u},\text{sp}} - T^{\text{a}}}{R^{\text{wua},\text{off}}}  + \frac{T^{\text{u},\text{sp}}-T^{\text{l},\text{sp}}}{R^{\text{ww}}}, & \text{if} \ \text{the lid is on}   \\
        \frac{T^{\text{u},\text{sp}} - T^{\text{a}}}{R^{\text{wua},\text{on}}} + \frac{T^{\text{u},\text{sp}}-T^{\text{l},\text{sp}}}{R^{\text{ww}}},   & \text{if} \ \text{the lid is off,} \\
    \end{cases}
\end{equation}
where $T^{\text{l},\text{sp}}$ and $T^{\text{u},\text{sp}}$ are the pre-specified setpoints in the lower and upper zones, respectively. In addition, $p^{\text{l},\text{Base}}$ is the steady-state power consumption for the lower zone, whereas $p^{\text{u},\text{Base}}$ is the steady-state power consumption for the upper zone, which takes two values depending of whether the lid is on or off.
Appendix B provides a simulation of (\ref{eq:StateSpaceModel}) using (\ref{eq1:steady-state-power}) and (\ref{eq2:steady-state-power}). The purpose of the steady-state operation is to highlight the benefit of flexibility provision from a more predictable, operational baseline consumption while the intricacies of the control logic is not investigated further here.
Furthermore, other benefits of using a  continuous power control include more optimal use of the heat in the furnace by being able to integrate the dipping schedule to the power consumption. Hence, smarter feed-forward planning of the lid can be achieved by utilizing continuous power control.

Early indications from the owner of the zinc furnace show a one-time cost of 0.5 million DKK into equipment and power electronics that allows for granular power control.

\vspace{1mm}
\section{Formulation for service provision} 
\label{opt}
\vspace{-1mm}
\subsection{Background and timeline for bidding decision making}
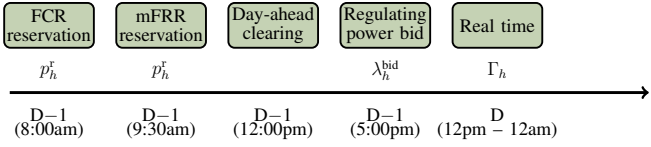
\begin{figure}
    \begin{adjustbox}{width=\columnwidth}
        \begin{tikzpicture}[scale=0.50,every text node part/.style={align=center}]

  \draw[rounded corners, fill=RectangleColor, very thick] (-5, 1) rectangle (-1, 3) node[midway]{\large FCR \\ \large reservation};
  \node at (-3,0) {\large  $p_{h}^{\rm{r}}$};
  \node at (-3,-2) {\large D$-$1};
  \node at (-3,-2.7) {\large (8:00am)};
  
  \draw[rounded corners, fill=RectangleColor, very thick] (0, 1) rectangle (4, 3) node[midway]{\large mFRR \\ \large reservation};
  \node at (2,0) {\large  $p_{h}^{\rm{r}}$};
  \node at (2,-2) {\large D$-$1};
  \node at (2,-2.7) {\large (9:30am)};

  \draw[rounded corners, fill=RectangleColor, very thick] (5, 1) rectangle (9, 3) node[midway]{\large Day-ahead \\ \large clearing};
  \node at (7,-2) {\large D$-$1};
  \node at (7,-2.7) {\large (12:00pm)};

  \draw[rounded corners, fill=RectangleColor, very thick] (10, 1) rectangle (14, 3) node[midway]{\large Regulating  \\ \large power bid};
  \node at (12,0) { \large $\lambda_{h}^{\text{bid}}$};
  \node at (12,-2) {\large D$-$1};
  \node at (12,-2.7) {\large (5:00pm)};

  \draw[rounded corners, fill=RectangleColor, very thick] (15, 1) rectangle (19, 3) node[midway]{\large Real time};
  \node at (17,0) {\large $\Gamma_{h}$};
  \node at (17,-2) {\large D};
  \node at (17,-2.7) {\large (12pm -- 12am)};

  \node (source) at (0-5, -1) {};
  \node (destination) at (24,-1){};
  \draw[->, line width=0.65mm, black](source)--(destination);

\end{tikzpicture}
    \end{adjustbox}
    \caption{\small{Decision making timeline and variables for bidding into FCR and mFRR markets of day $\rm{D}$. 
    \vspace{-3mm}}}
    \label{fig:timeline}
\end{figure}

\vspace{-1mm}
FCR is the fastest responding ancillary service in the Danish bidding zone of DK1, which is the western part of Denmark and shares the same frequency as continental Europe. The response is automatic and must be adjusted according to frequency deviations up to $\pm 200$ mHz around the nominal frequency of 50 Hz \cite{energinet:prequalification}. The service is currently delivered primarily by thermal power plants and batteries, but flexible demands can potentially also deliver FCR, either by control of individual assets or from an aggregation of multiple assets that can be turned on/off in a manner that adheres to the frequency response requirements. The service is bought by the Danish TSO, Energinet, in 4-hour blocks in one auction for the upcoming day as shown in Fig. \ref{fig:timeline}. The auction occurs at 8:00am the day before delivery. FCR providers receive the marginal price of the auction as payment for providing reservation capacity.

mFRR is the last service activated by the TSO manually after a frequency drop, such that the amount of energy injected to the grid can be several MW. In DK1, Energinet procures around 600 MW of mFRR for every hour of operation \cite{energinet:scenario_report_2022}. In Denmark, there is only a market for upward mFRR, i.e., when the frequency drops. 
As described in \cite{gade2023load} and shown in Fig. \ref{fig:timeline}, the procurement of mFRR reserves happens before the day-ahead (spot) market clearing at 9:30am and mFRR providers are paid at the marginal price according to the capacity they offer. At 5:00pm of the day before, the mFRR providers submit their price bids, so-called regulating power bids for activation. 
In real-time when the TSO requires, given regulating power bids, activation of reserves happens starting from the mFRR provider with the lowest regulating power bid. This leads to a balancing price higher than the day-ahead price. mFRR providers are paid according to the amount they activate and penalized if they do not deliver their promised capacity. For TCLs such as zinc furnaces, there will be an inevitable rebound which requires additional energy.

In the following we present an optimization model for offering FCR services, and then another optimization model to offer mFRR services.
Both models are linear and deterministic with full hindsight on FCR/mFRR and spot electricity prices and frequency deviations. The online appendix \cite{code} provides historical prices in DK1 from January 2021 to July 2023 for FCR, mFRR, and spot electricity prices. By such a perfect foresight assumption, we provide an upper bound for the profit of flexibility potential which is useful to know when deciding to make investments enabling power flexibility provision. If the upper bound potential is unattractive, then investments should not be made.

\vspace{1mm}
\subsection{FCR}
\vspace{-1mm}
The linear optimization model for offering FCR reads as
%
\begin{subequations}\label{P1:compact_model_fcr}
    \begin{align}
        \underset{\bm{\Gamma}_{h},\bm{\Gamma}_{t}}{\textrm{max}} \quad & \bm{p}^{r} \bm{\lambda}^{\text{FCR}} - \bm{s}\bm{\lambda}^{\text{Pen}} \label{P1:eq1}
        \\
        s.t \quad                                                      & q(\bm{\Gamma}_{h}) \leq 0 \label{P1:eq2}                                                                                                                      \\
        \quad                                                          & \text{State-space model in } (\ref{eq:StateSpaceModel}), \label{P1:eq8}                                                                                       \\
        \quad
                                                                       & \bm{\Gamma}_{h} = \Bigl( \bm{p}^{r}, \bm{s} \Bigr) \in \mathbb{R}  \label{P1:eq9}                                                                             \\
        \quad
                                                                       & \bm{\Gamma}_{t} = \Bigl( \bm{p}, \bm{p}^{q}, \bm{s}^{q, \prime}, \bm{T}^{\text{z},q}, \bm{T}^{\text{w},q} \Bigr) \in \mathbb{R}, \ \forall{q}, \label{P1:eq10}
    \end{align}
\end{subequations}
where $h$ and $t$ are index for hours and minutes, respectively. Bold symbols indicate time vectors. The objective function (\ref{P1:eq1}) maximizes the revenues (first term) minus penalties for failures (second term). The variable vector $\bm{p}^{r}$ includes hourly FCR capacity bids (in kW), while the hourly FCR prices $\bm{\lambda}^{\text{FCR}}$ (in DKK/kW) are given. The flexibility provider incurs a penalty cost at rate $\bm{\lambda}^{\text{Pen}}$ (in DKK/kWh) when fails in delivering the promised service, measured by hourly variables $\bm{s}$ (in kW).
Variables related to hourly bidding capacity and penalties are included in $\bm{\Gamma}_{h}$ as defined in \eqref{P1:eq9}. Similarly, variables related to real-time control, indexed by minute $t$, are within $\bm{\Gamma}_{t}$ as defined in \eqref{P1:eq10}.
The inequality constraints in (\ref{P1:eq2}) are presented in (\ref{P1:constraints-1}) and (\ref{P1:constraints-2}).
The superscript $q$ denotes the two zones $\{l, u\}$ in the zinc furnace, and the total reserve from the zinc furnace is simply their summation. Note that $\bm{s}^{q}$ is the hourly variable vector indicating the amount of failure in the corresponding zone $q$ (either upper or lower zone), whereas $\bm{s}^{q, \prime}$ is similar but in the minute resolution. The set of constraints in (\ref{P1:eq2}) related to frequency response are:
\begingroup
\allowdisplaybreaks
\begin{align}
    \bm{p}^{q} = \bm{F} \bm{p}^{q,r} + \bm{s}^{q, \prime} + \bm{P}^{q, \text{Base}}
    \quad \bm{s}^{q} \geq |\bm{s}^{q, \prime}|,
    \quad \forall{q}  \label{P1:constraints-1}
\end{align}
\endgroup
where the parameter $\bm{F}$ is normalized between -1 and 1 depending on the frequency. This parameter represents the normalized response required \cite{energinet:prequalification} as
\begingroup
\allowdisplaybreaks
\begin{equation}
    \begin{cases}
        -1,                             & \text{if}\ F_{t} \leq 49.8\ \text{Hz}                                          \\
        \frac{F_{t}-50+0.02}{0.2-0.02}, & \text{if}\ F_{t} \leq 49.98\ \text{Hz}\ \text{and}\ F_{t} \geq 49.8\ \text{Hz} \\
        \frac{F_{t}-50-0.02}{0.2-0.02}, & \text{if}\ F_{t} \leq 50.2\ \text{Hz}\ \text{and}\ F_{t} \geq 50.02\ \text{Hz} \\
        1,                              & \text{if}\ F_{t} \geq 50.2\ \text{Hz}.                                         \\
    \end{cases}
\end{equation}
\endgroup

The remaining constraints in (\ref{P1:eq2}) are related to total power and violation from both zones in the furnace:
\begingroup
\allowdisplaybreaks
\begin{subequations} \label{P1:constraints-2}
    \begin{align}
        \quad & \bm{p}^{q,r} \leq \bm{P}^{q,\text{Base}} , \quad \forall{q} \label{P1:co8}
        \\
        \quad & \bm{p} = \sum_{q} \bm{p}^{q},
        \quad \bm{p}^{r} = \sum_{q} \bm{p}^{q,r},
        \quad \bm{s} = dt \sum_{q} \bm{s}^{q}.   \label{P1:co7}
    \end{align}
\end{subequations}
\endgroup

\vspace{-1mm}
\subsection{mFRR}
\vspace{-1mm}
The optimization model for offering mFRR services is a mixed-integer linear program, identical to that in \cite{gade2023load}, which is the earlier publication of the authors. The only difference is that we consider the state-space model as represented in  (\ref{eq:StateSpaceModel}) and a deterministic framework. In a compact form, this  optimization model reads as
%
%
\begin{subequations}\label{P1:compact_model_mfrr}
    \begin{align}
        \underset{\bm{p}^{\rm{r}}, \bm{p}^{q,\rm{r}}, \bm{\lambda}^{\rm{bid}}, \bm{\Gamma}}{\text{Maximize}} \  & f(\bm{p}^{\rm{r}}, \bm{p}^{q,\rm{r}}) + g(\bm{\Gamma}) \label{P1:eq1_mfrr}
        \\
        \text{s.t.} \                                                                                           & h(\bm{p}^{\rm{r}}, \bm{p}^{q,\rm{r}}, \bm{\lambda}^{\rm{bid}}, \bm{\Gamma}) \leq 0, \label{P1:eq2_mfrr}                                              \\
        \                                                                                                       & \text{State-space model } (\ref{eq:StateSpaceModel}),  \label{P1:eq3_mfrr}
        \\
        \                                                                                                       & \bm{T}^{\text{z},q}, \bm{T}^{\text{w},q} \in \mathbb{R}, \forall{q} \label{P1:eq4_mfrr}
        \\
        \                                                                                                       & \bm{T}^{\text{z},q,\text{Base}}, \bm{T}^{\text{w},q,\text{Base}}, \in \mathbb{R}, \ \forall{q} \label{P1:eq5_mfrr}
        \\
        \                                                                                                       & \Bigl( \bm{p}, \bm{p}^{q}, \bm{p}^{\rm{r}}, \bm{p}^{q,\rm{r}}, \bm{\lambda}^{\rm{bid}}, \bm{p}^{\rm{b},\uparrow}, \bm{p}^{\rm{b},\downarrow}, \notag \\ & \bm{p}^{q,\rm{b},\uparrow}, \bm{p}^{q,\rm{b},\downarrow}, \bm{s}, \bm{s}^{q}, \bm{\phi} \Bigr) \in \mathbb{R}^{+}, \ \forall{q}  \label{P1:eq6_mfrr}
        \\
        \                                                                                                       & \Bigl( \bm{g}, \bm{u}^{q,\uparrow}, \bm{z}^{q,\uparrow}, \bm{y}^{q,\uparrow}, \notag                                                                 \\ & \bm{u}^{q,\downarrow}, \bm{z}^{q,\downarrow}, \bm{y}^{q,\downarrow} \Bigr) \in \{0,1\}, \ \forall{q}, \label{P1:eq7_mfrr}
    \end{align}
\end{subequations}
which determines the optimal hourly mFRR capacity bids $\bm{p}^{r}$ and optimal hourly regulating price bids $\bm{\lambda}^{\text{bid}}$. The full model formulation is available in the online appendix \cite{code}, where all source codes are also publicly available.


\vspace{2mm}
\section{Numerical Results}
\label{results}
\vspace{-1mm}
\subsection{Cost saving analysis}
\begin{figure}[t]
    \centering
    \includegraphics[width=0.75\columnwidth]{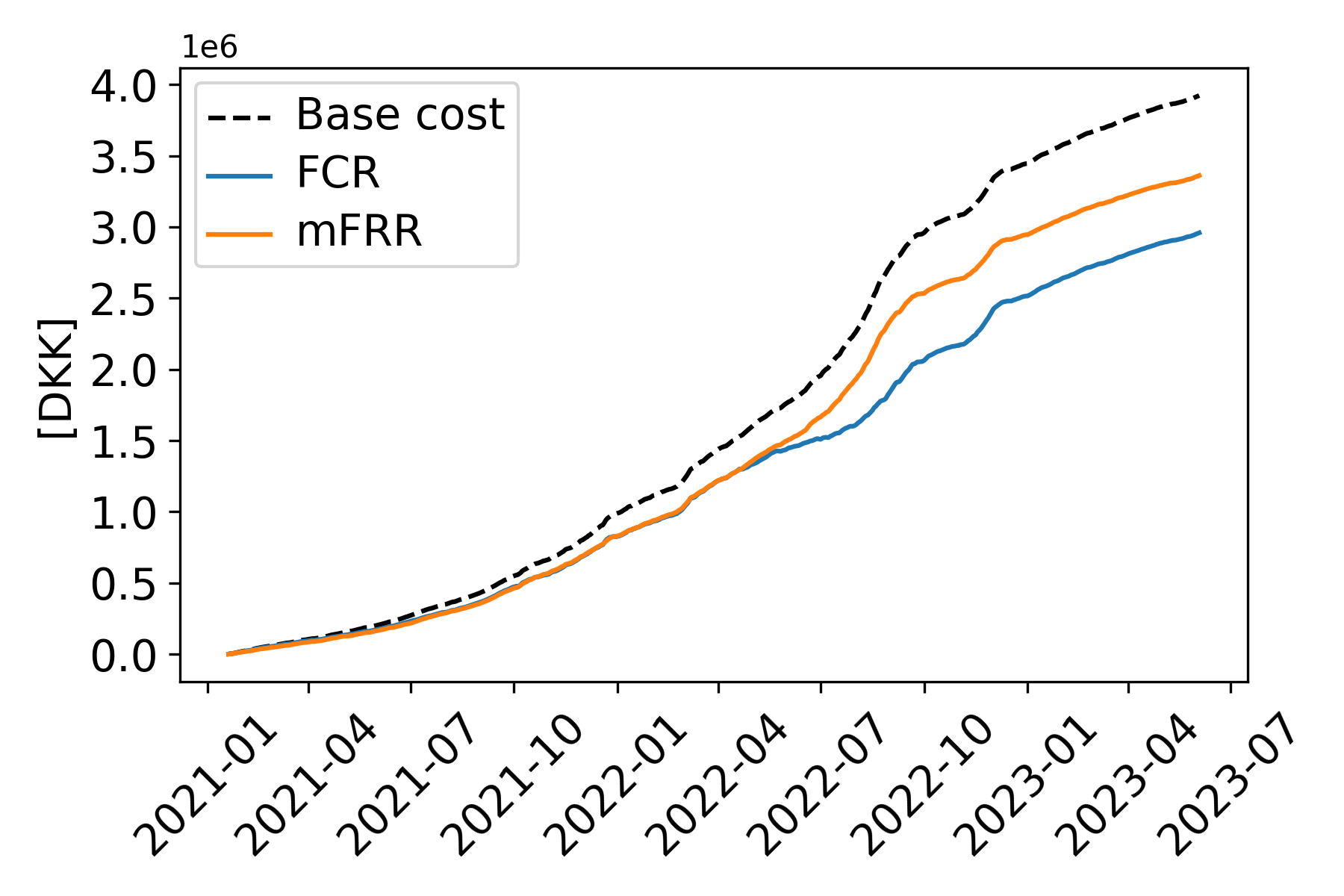}
    \caption{\small{Cumulative operational cost of the zinc furnace in the period of January 2021 to July 2023 when participating in either FCR or mFRR market compared to the baseline (base cost) when no services are offered. \vspace{-3mm}}}
    \label{fig:cumulative_cost_comparison}
\end{figure}

\vspace{-1mm}
Fig. \ref{fig:cumulative_cost_comparison} shows the cumulative operational cost of the zinc furnace in the period of January 2021 to July 2023 when participating in either FCR or mFRR market compared to the baseline (base cost) when no services are provided. Clearly, FCR provides a remarkable cost saving during the summer when prices are high. Cost savings for FCR compared to mFRR seem to be mostly due to this as FCR stagnates a bit after September 2022, when FCR bidding was open to all of Europe and not just the western Denmark.

Although Fig. \ref{fig:cumulative_cost_comparison} shows an upper bound for the profit potential, there is still a significant benefit for a zinc furnace to offer its flexibility. A saving of around 1 million DKK over a two and a half year period can quickly recover the one-time investment cost of 0.5 million DKK, needed to enable smart control of the power supply to the furnace. However, savings were not impressive in 2021 and 2023 when FCR prices were comparatively low. The Danish TSO expects more balancing demand in the future, and therefore FCR should remain an attractive revenue stream \cite{energinet:scenario_report_2022}.

\vspace{1mm}
\subsection{Two worst days}
\vspace{-1mm}
To investigate how the flexibility provision differs for FCR and mFRR, two worst days in the period of January 2021 to July 2023 are extracted and analyzed.

\begin{figure}[t]
    \centering
    \includegraphics[width=\columnwidth]{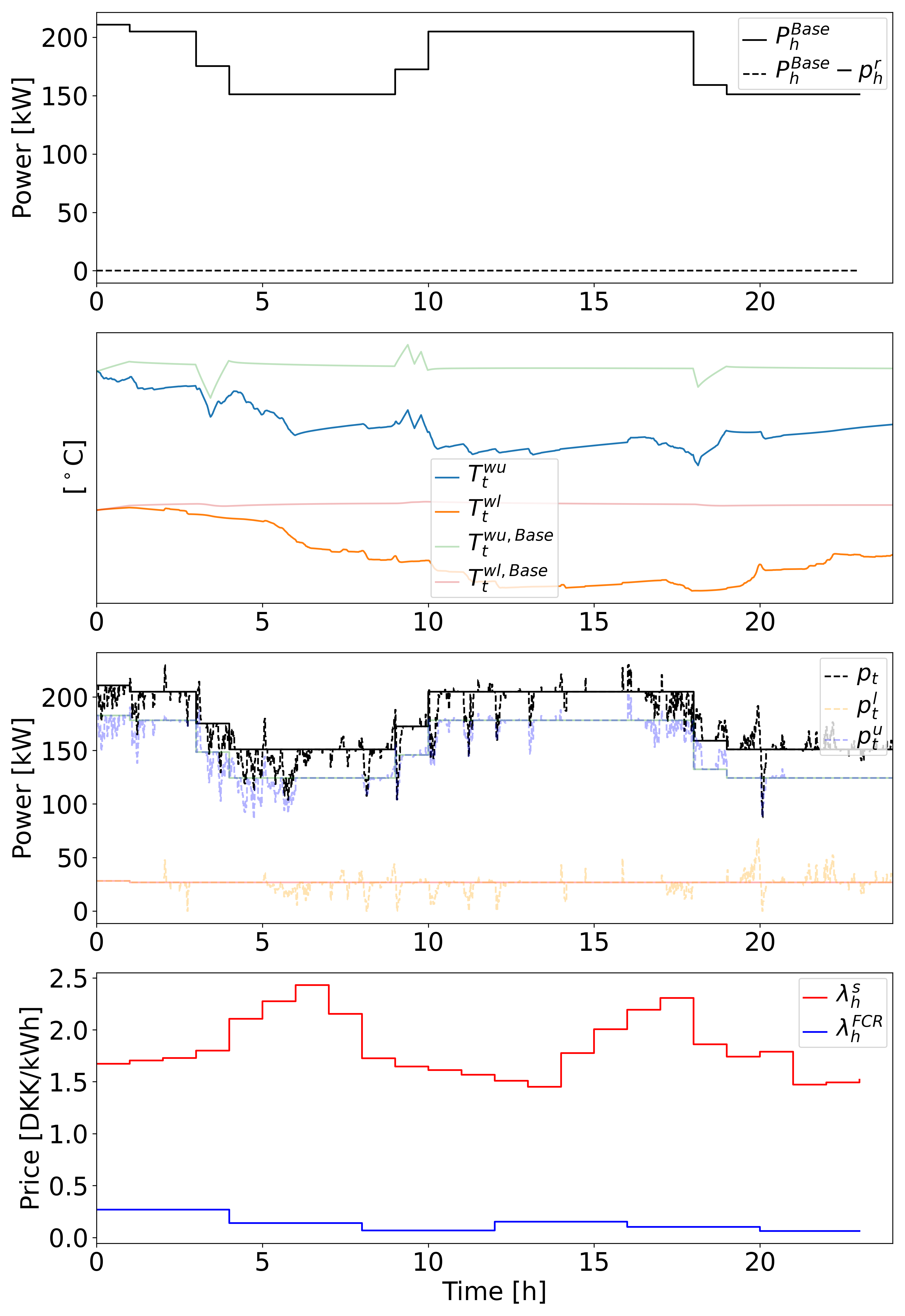}
    \caption{\small{Worst-case day in the period of January 2021 to July 2023 in terms of cumulative frequency deviation to 50 Hz. \textbf{Top}: total baseline operational power and FCR reserve capacity. \textbf{Middle upper}: wall temperature of upper and lower zones with and without (denoted as base) FCR market participation. \textbf{Middle lower}: operational power consumption of lower and upper zones as well as the total consumption with the baseline operational consumption in solid lines. \textbf{Bottom}: spot and FCR prices for the worst-case day. \vspace{-3mm}}}
    \label{fig:fcr_single_case}
\end{figure}

For FCR, the worst day is defined as the day where the frequency deviated the most from 50 Hz. The operational results for this day are shown in Fig. \ref{fig:fcr_single_case}. The entire baseline consumption is offered as a reserve service (top plot). This happens in all days  as it maximizes the revenue from FCR. The implication is that the furnace has to provide a frequency response \textit{continuously} for all hours  which has an impact on the temperature (middle upper plot). The temperature  declines slightly and is still satisfactory within the thresholds before solidification occurs. Due to the anonymization, thresholds and the y-axis scale are not shown. The frequency response (middle lower plot) shows that the response is mostly up-regulation, i.e., turning down the power consumption, which happens when the frequency is below 50 Hz.

\begin{figure}[t]
    \centering
    \includegraphics[width=\columnwidth]{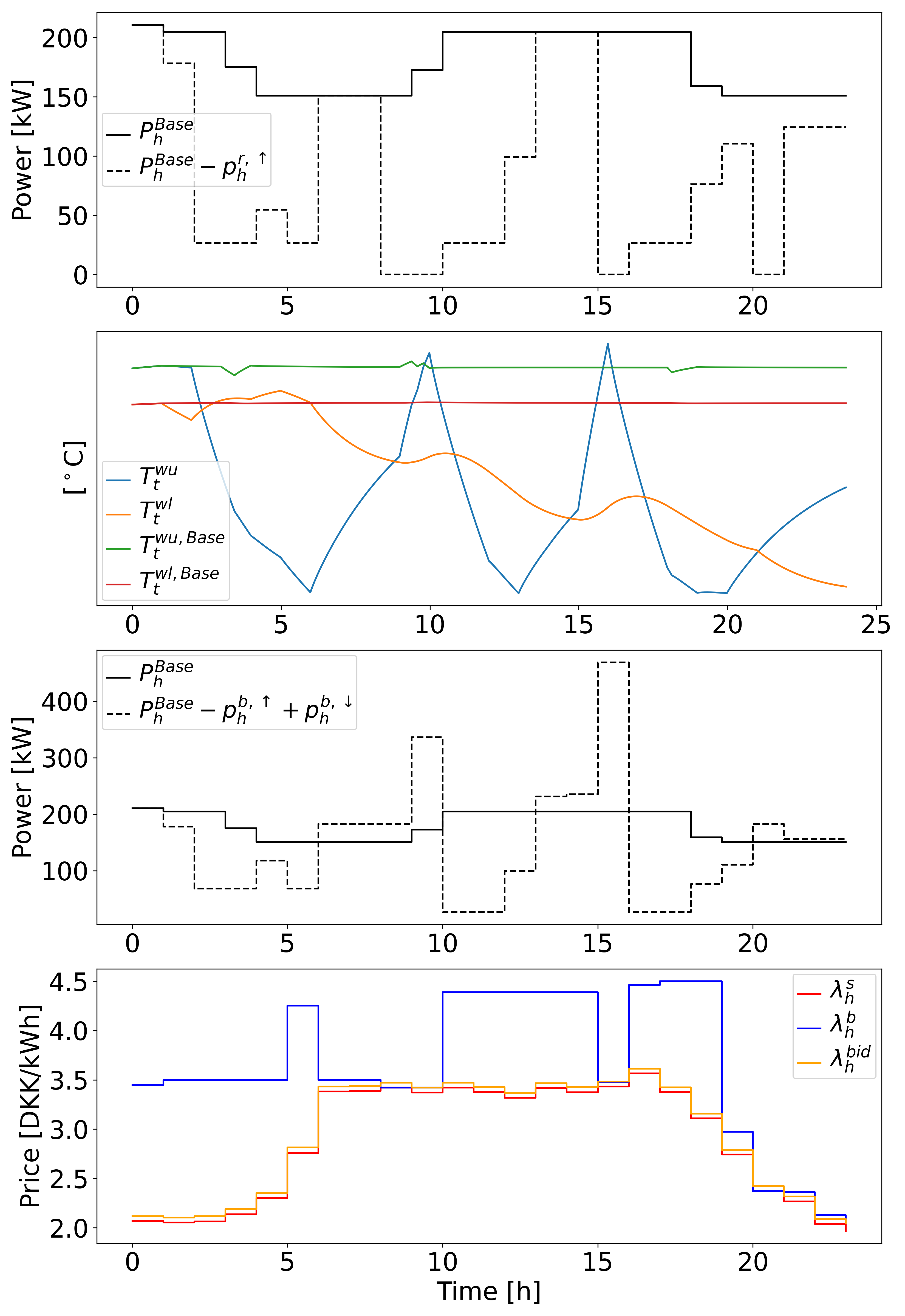}
    \caption{\small{Worst-case day in the period of January 2021 to July 2023 in terms of up-regulation in the power grid. The captions for first three plots are the same as those in Fig. \ref{fig:fcr_single_case} but for mFRR. The bottom plot shows the spot and balancing prices for the worst-case day together with regulating power price bids. \vspace{-3mm}}}
    \label{fig:mfrr_single_case}
\end{figure}

For mFRR, the reserve quantities are slightly lower than that for FCR as observed in Fig. \ref{fig:mfrr_single_case} (top). This is due the penalty of not being able to fully deliver  up-regulation as it can be seen by comparing the bottom plot with the reserve (top) and actual consumption (middle lower): whenever the bid is lower than the balancing price \textit{and} a reserve is promised, then the actual power consumption should correspond to $P^{\text{Base}}_{h} - p_{h}^{r, \uparrow}$. For example, this does not happen in hour 12. Also, the optimization model in (\ref{P1:compact_model_mfrr}) requires an immediate rebound after activation which prohibits the zinc furnace from up-regulating more than 12 hours per day. For these reasons, the reserved power is not equal to the operational baseline power.
Furthermore, turning off the power consumption for five consecutive hours  affects the temperature as observed in the middle upper plot. Here, the impact is much bigger than that for FCR and exceeds the pre-specified thresholds by an order of magnitude.

\begin{figure}[t]
    \centering
    \begin{subfigure}{0.49\columnwidth}
        \centering
        \includegraphics[width=0.98\columnwidth]{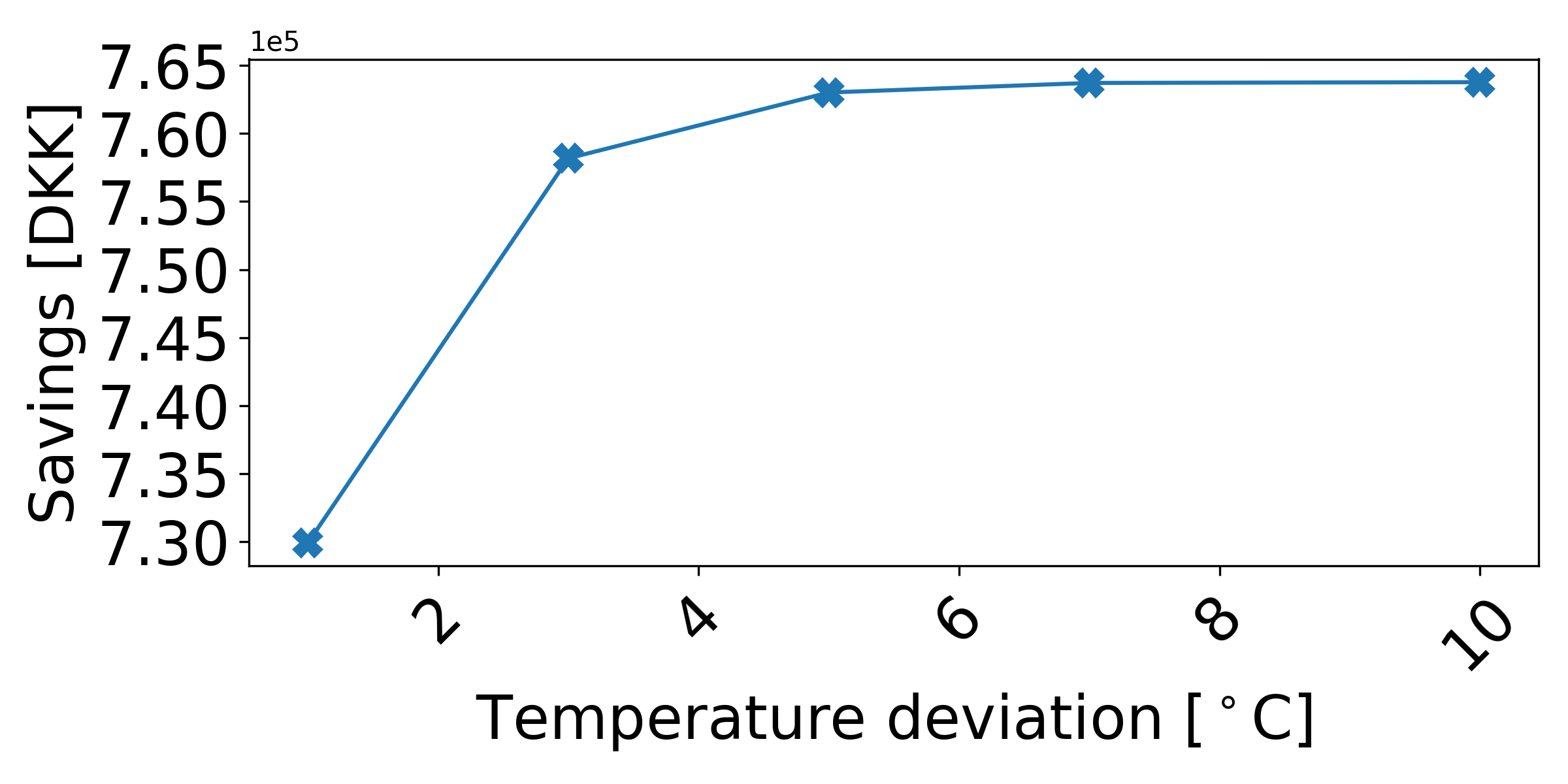}
        \caption{Monetary savings for FCR}
        \label{subfig:profit_vs_delta_temp_fcr}
    \end{subfigure}
    \begin{subfigure}{0.49\columnwidth}
        \centering
        \includegraphics[width=0.98\columnwidth]{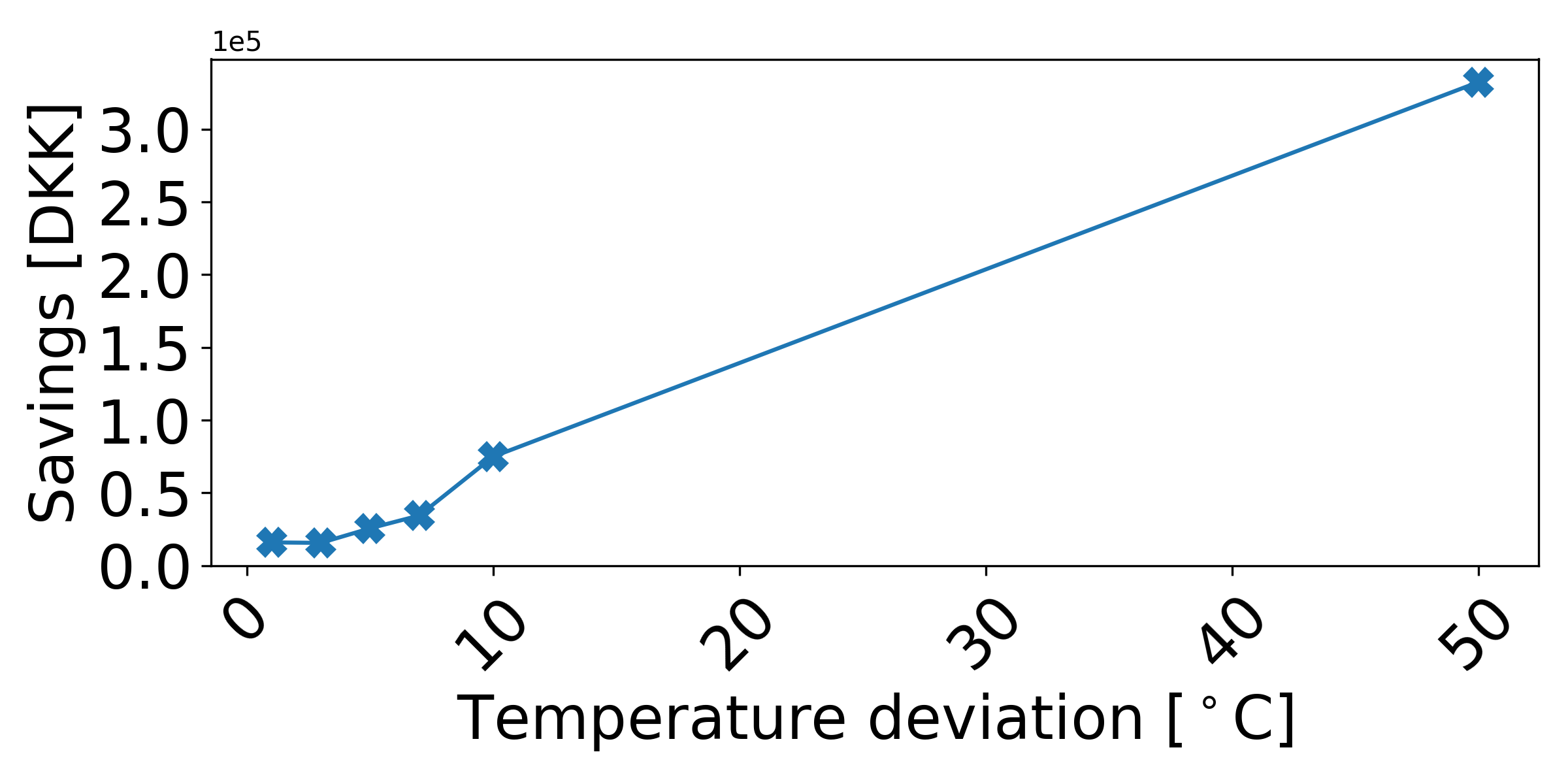}
        \caption{Monetary savings for mFRR}
        \label{subfig:profit_vs_delta_temp_mfrr}
    \end{subfigure}
    \caption{\small{Monetary savings as a function of constraining the allowed temperature deviation of the zinc to the setpoint. \vspace{-3mm}}}
    \label{fig:profit_vs_delta_temp}
\end{figure}

\vspace{1mm}
\subsection{Constraining temperature deviations}
\vspace{-1mm}
To assess the available flexibility in the zinc furnace, it is prudent to investigate the temperature deviation impact on the monetary savings for both FCR and mFRR services. For optimization models in (\ref{P1:compact_model_fcr}) and (\ref{P1:compact_model_mfrr}), additional constraints $ \Delta^{\text{min}} \leq \bm{T^{w,q}} \leq \Delta^{\text{max}} \ \forall{q}$ are added to constrain the allowed temperature deviation.
As already indicated  in Fig. \ref{fig:fcr_single_case}, the temperature is unaffected to any significant degree for FCR. This is again shown in Fig. \ref{subfig:profit_vs_delta_temp_fcr}, where an allowed temperature deviation of 1 C$^{\circ}$ is enough to provide substantial monetary savings. This is not the case for mFRR as observed in Fig. \ref{subfig:profit_vs_delta_temp_mfrr} and alluded to previously in Fig. \ref{fig:mfrr_single_case}. Fig. \ref{subfig:profit_vs_delta_temp_mfrr} shows diminishing value at lower temperature deviations while the most savings are obtained when allowing temperature deviations of at least 6 C$^{\circ}$.




\vspace{1mm}
\section{Conclusion} \label{conc}
\vspace{-1mm}
This paper explored how a single-state industry process as exemplified by a real-world case study of a zinc galvanizing process can make modest investments to enable flexibility provision in terms of FCR and mFRR services. By switching static control logic to continuous power control, economic benefits can be achieved. Comparing FCR and mFRR, it is certainly not realistic to expect a full up-regulation of five consecutive hours for mFRR for a zinc furnace, which will not happen when delivering FCR as observed in two worst days analyzed. Besides, FCR is more profitable over the time period we considered. The FCR provision looks a more attractive opportunity for a zinc furnace owner as it provides a stable and passive source of income once investments into continuous power control are made.

The revenue earned from FCR and mFRR provision is typically shared between the aggregator and flexible consumer. Hence, the upper bound savings reported in this paper are also bound to a payment agreement with the aggregator. That can potentially make it less attractive for a flexible demand to participate in ancillary service markets. However, the FCR market still seems like an obvious option and likewise for the investments made to enable the FCR provision. mFRR, however, is not too attractive as it can have a detrimental impact on the zinc temperature on days with severe up-regulation. Although less profitable, mFRR strategies can be employed. For example, the provision of real-time balancing only without reserves is a viable option since no commitments are made beforehand.

Other revenue streams such as load shifting and automatic frequency restoration reserve (aFRR) services were not considered in this paper. However, they can also potentially  be profitable in certain market regimes with a high aFRR demand or very volatile spot prices as experienced in 2022.

{\appendices

\vspace{1mm}
\section*{Appendix A: Parameter estimation}\label{app:parameter-estimation}
\vspace{-1mm}
The parameters and latent states in (\ref{eq:StateSpaceModel}) have been estimated using CTSM-R \cite{juhl2016ctsmr} as implemented in \cite{code}. The one-step residuals in the estimation procedure are shown in Fig. \ref{fig:4thOrderModelValidation} and resembles white noise (middle plot). Furthermore, the autocorrelation shows no significant lags, and the cumulative periodogram illustrates no frequencies with significant power. Hence, the model in (\ref{eq:StateSpaceModel}) captures the temperature dynamics well for both  upper and lower zones of the furnace.

\begin{figure}[t]
    \centering
    \includegraphics[width=\columnwidth]{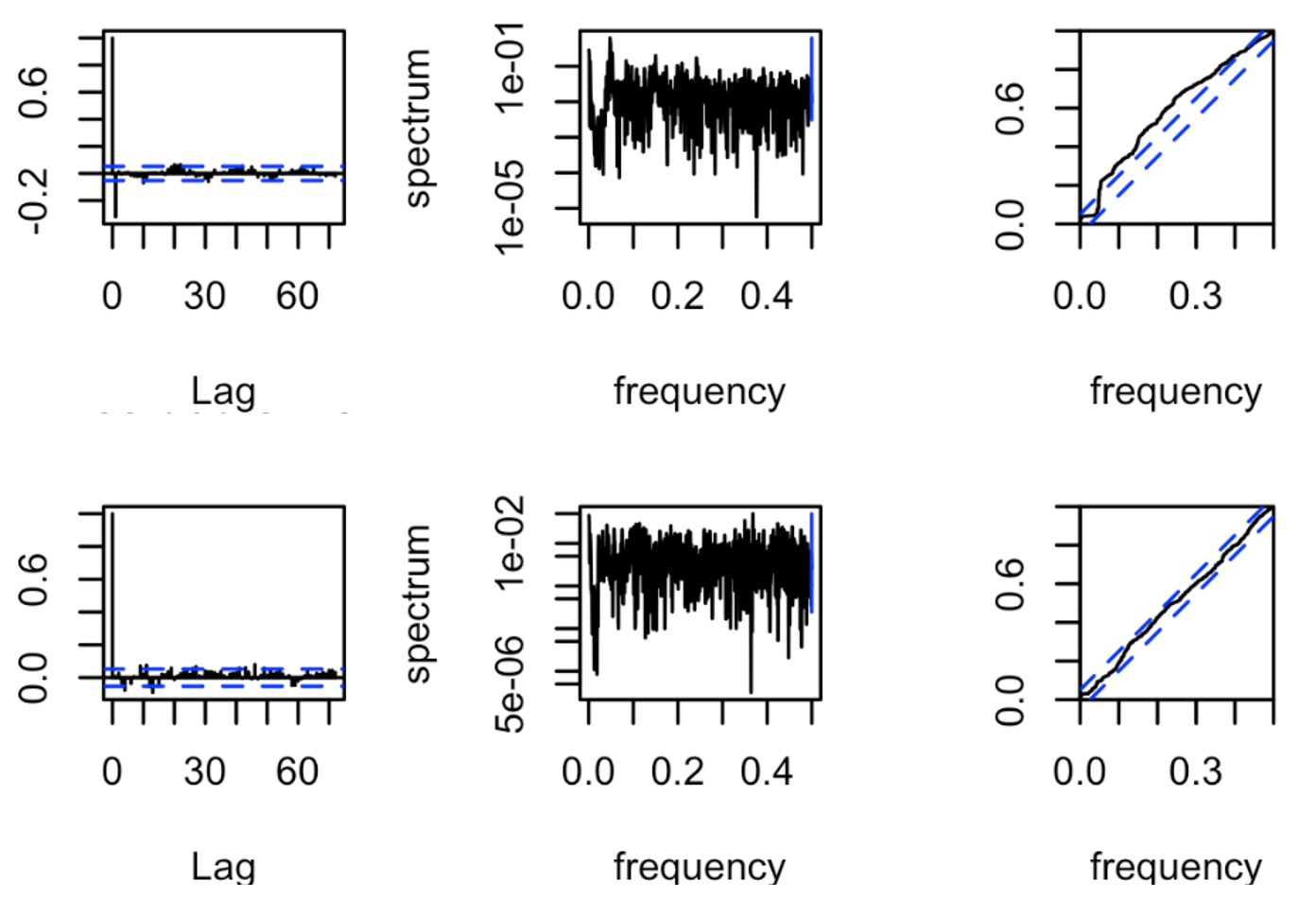}
    \caption{\small{Validation of state-space model in (\ref{eq:StateSpaceModel}). \textbf{Top}: upper zone wall temperature. \textbf{Bottom}: lower zone wall temperature. \textbf{Left}: autocorrelations of the model residuals. \textbf{Middle}: residuals. \textbf{Right}: cumulative periodogram of the residuals. \vspace{-3mm}}}
    \label{fig:4thOrderModelValidation}
\end{figure}

\vspace{1mm}
\section*{Appendix B: Steady-state Simulation}\label{app:steady-state-simulation}
\vspace{-1mm}
Fig. \ref{fig:4thOrderModelVisualizationSteadyState} shows a 24-hour simulation (1440 time steps) of (\ref{eq:StateSpaceModel}) using (\ref{eq1:steady-state-power}) and (\ref{eq2:steady-state-power}), which was chosen to include both regimes where the lid is either on or off. The original data is shown in dashed black lines. It is evident how the steady-state power consumptions in (\ref{eq1:steady-state-power}) and (\ref{eq2:steady-state-power}) and temperatures are  more stable and predictable. To achieve this, a controller is needed to keep the temperature at the setpoint.

\begin{figure}[t]
    \centering
    \includegraphics[width=\columnwidth]{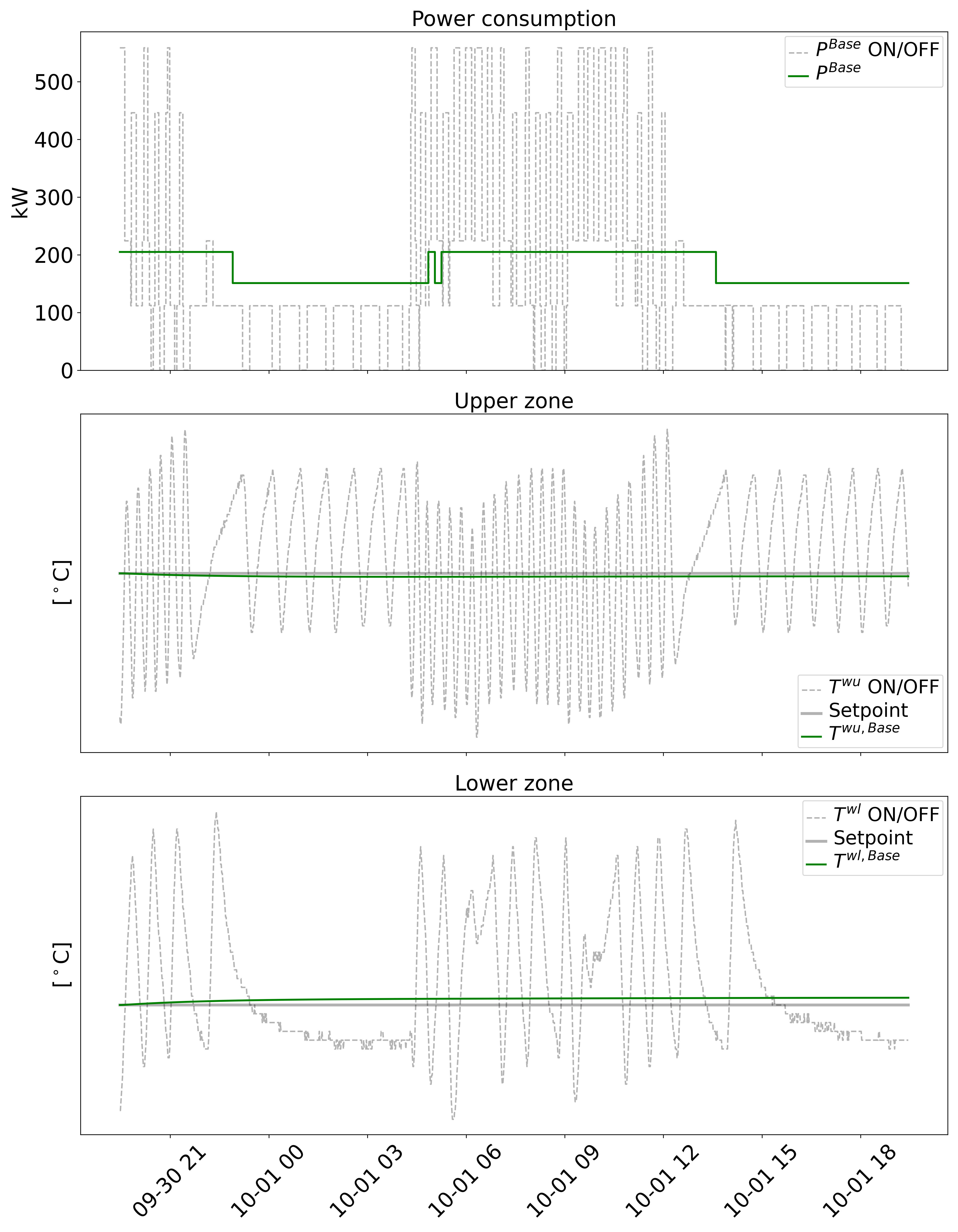}
    \caption{\small{Simulation of (\ref{eq:StateSpaceModel}) with $p^{\text{u}}$ and $p^{\text{l}}$ set to the steady-state consumptions as specified in (\ref{eq1:steady-state-power}) and (\ref{eq2:steady-state-power}). \textbf{Top}: the total power consumption in steady-state and original data (with ON/OFF control). \textbf{Middle}: the upper zone wall temperature at steady-state power consumption and original data (with ON/OFF control). \textbf{Bottom}: the lower zone wall temperature at steady-state power consumption and original data (with ON/OFF control). \vspace{-3mm}}}
    \label{fig:4thOrderModelVisualizationSteadyState}
\end{figure}


}


\vspace{-1mm}
\bibliographystyle{IEEEtran}
\bibliography{tex/bibliography/Bibliography}


\newpage

\section*{Online Appendix A: Market Prices}\label{app:market-prices}

\begin{figure}[t]
    \centering
    \includegraphics[width=\columnwidth]{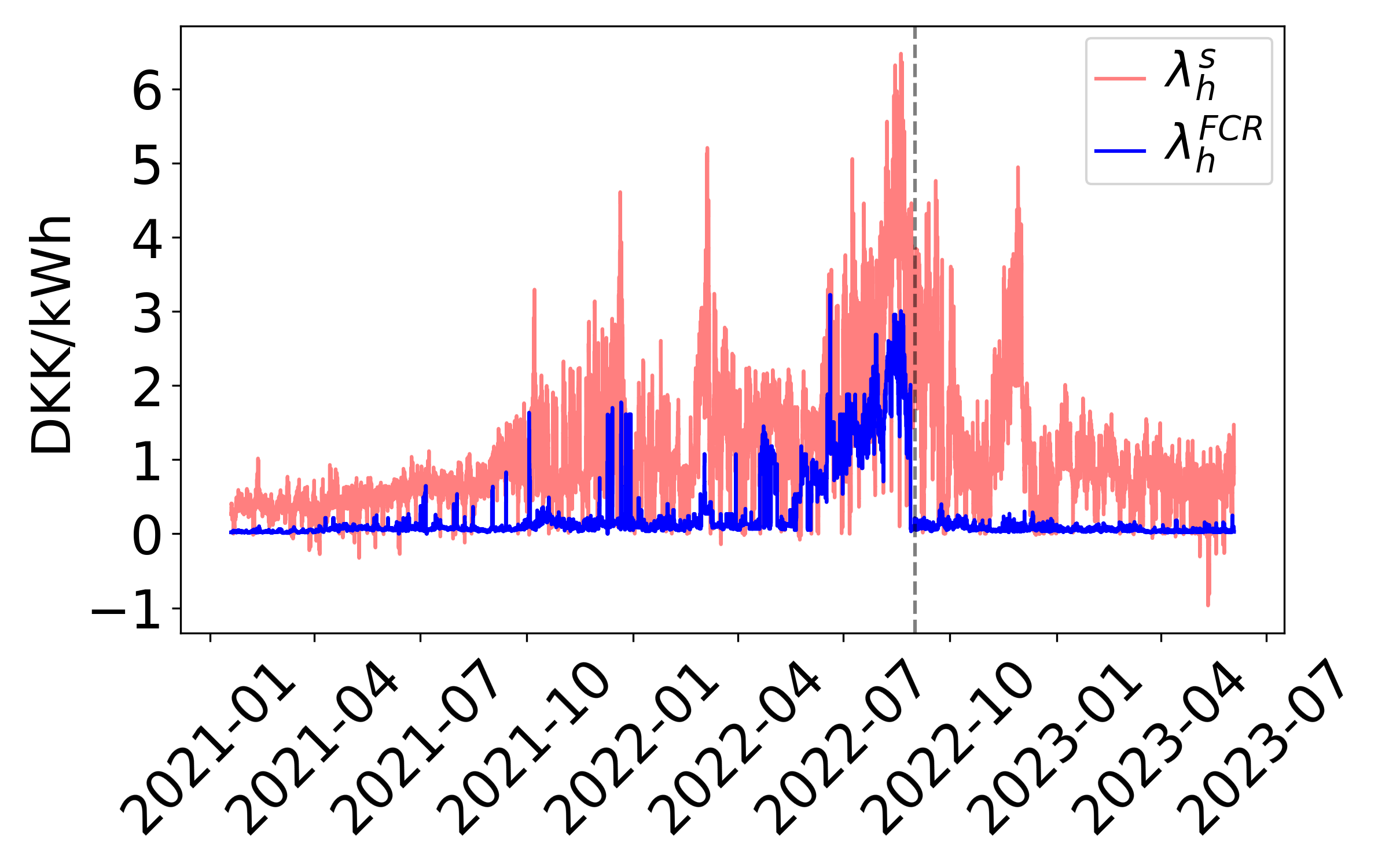}
    \caption{\small{FCR and spot prices in the period of January 2021 to July 2023 in the Danish bidding zone of DK1. In September 2022, FCR tenders included continental Europe.}}
    \label{fig:fcr_prices_2022}
\end{figure}

Fig. \ref{fig:fcr_prices_2022} shows the FCR ($\lambda_{h}^{\text{FCR}}$) and spot prices ($\lambda_{h}^{\text{s}}$) in DK1 in the period of January 2021 to July 2023. Notice how FCR prices went down after September 2022 where the Danish TSO opened up market participation from the entire continental Europe. Furthermore, FCR prices have been proportional to spot prices.

\begin{figure}[t]
    \centering
    \includegraphics[width=\columnwidth]{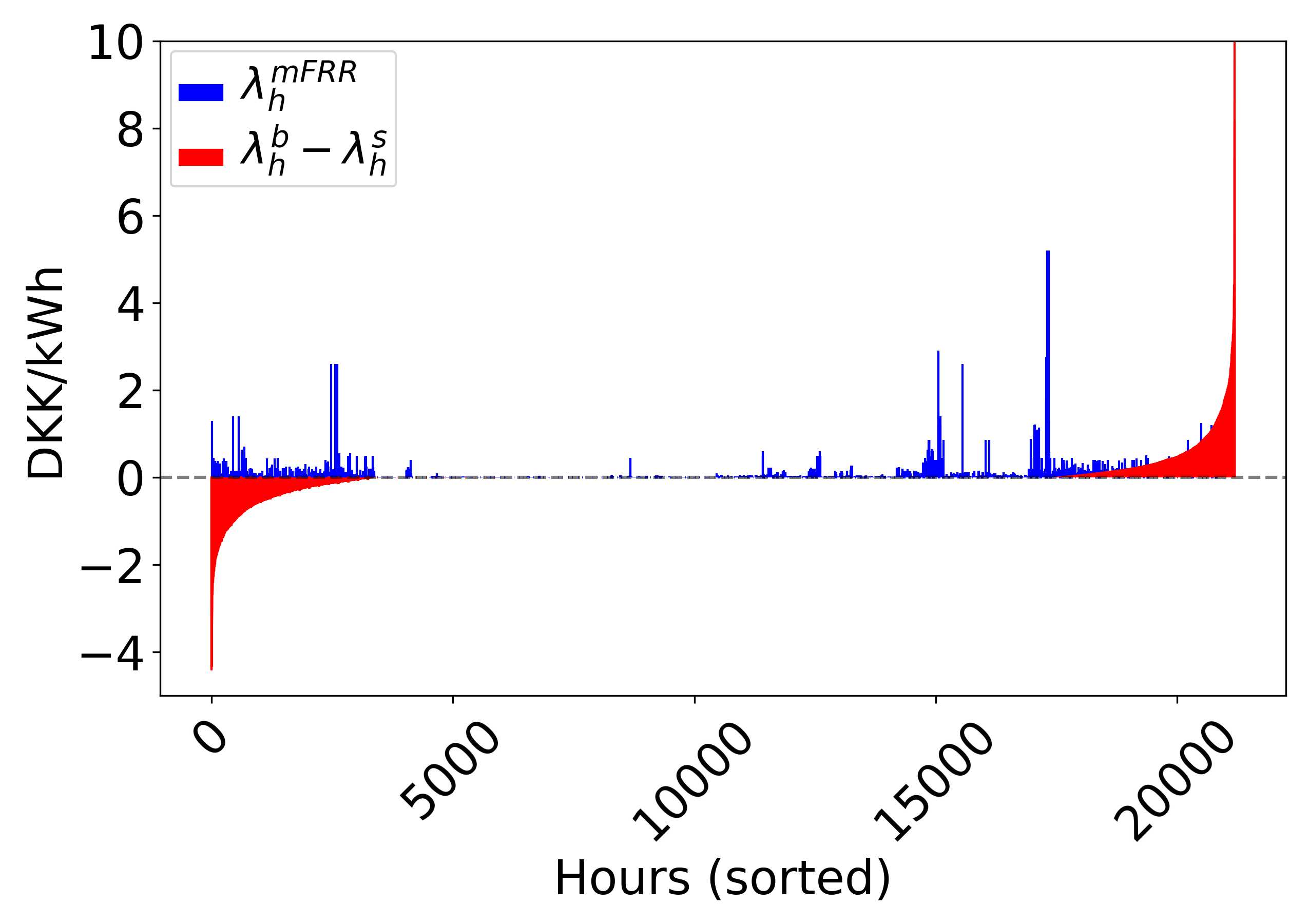}
    \caption{\small{Hourly mFRR prices and balancing price differentials in the time period of January 2021 to July 2023 in ascending order.}}
    \label{fig:mfrr_prices_2022}
\end{figure}

Fig. \ref{fig:mfrr_prices_2022} shows the distribution of hourly balancing prices ($\lambda_{h}^{\text{b}}$) minus spot prices (referred to as balance price differentials) ordered from low to high in red. The mFRR prices ($\lambda_{h}^{\text{mFRR}}$) in the same hours are shown in blue. When balance price differentials are below zero, down-regulation takes place, i.e., supply is greater than demand. When price differentials are above zero, up-regulation takes place, i.e., demand is higher than supply. The mFRR capacity is for up-regulation only, but flexible providers can both down- and up-regulate in the real-time balancing market if they choose. Here, however, we only consider the mFRR up-regulation capacity market where the zinc furnace is paid for both capacity as shown in blue in Fig. \ref{fig:mfrr_prices_2022} and actual up-regulations as shown in red when the balance price differential is above zero.

\section*{Online Appendix B: Mixed-integer linear optimization for offering mFRR services}\label{appendix:D}






\begingroup
\allowdisplaybreaks
The mixed-integer linear optimization problem for the zinc furnace to optimally bid in the mFRR market reads as
%
\begin{subequations}\label{P2:FinalModel}
    \begin{align}
           & \underset{\bm{p}^{\rm{r}}, \bm{p}^{q,\rm{r}}, \bm{\lambda}^{\rm{bid}}, \bm{\Gamma}}{\text{Maximize}} \ \sum_{h=1}^{24}\lambda_{h}^{\rm{r}} p^{\rm{r}}_{h}+ \Bigl(\sum_{h=1}^{24}  \lambda_{h}^{\rm{b}} p^{\rm{b},\uparrow}_{h} - \notag                                                                                                                                                                                                                                                            \\  &  \hspace{0.5cm}\sum_{h=1}^{24}  \lambda_{h}^{\rm{b}} p^{\rm{b},\downarrow}_{h} - \sum_{h=1}^{24}  \lambda^{\rm{p}}s_{h} \Bigr) \label{P2:1} \\
           & \   \text{s.t.}  \  (\ref{P1:constraints-2}), \ \forall{h},   \label{P2:2}                                                                                                                                                                                                                                                                                                                                                                                                                                                           \\
           & \ p^{\rm{b},\downarrow}_{h} = p^{\rm{zu}, \rm{b},\downarrow}_{h} + p^{\rm{zl}, \rm{b},\downarrow}_{h}, \ \forall{h} \label{P2:3} \\
           & \ p^{\rm{b},\uparrow}_{h} = p^{\rm{zu}, \rm{b},\uparrow}_{h} + p^{\rm{zl}, \rm{b},\uparrow}_{h}, \ \forall{h} \label{P2:33} \\
        T^{\text{zu}}_{t+1} & = T^{\text{zu}}_{t} + dt \cdot \frac{1}{C^{\text{zu}}}\Bigl( \frac{1}{R^{\text{zuzl}}} (T^{\text{zl}}_{t} - T^{\text{zu}}_{t}) \notag                                     \\ & \mspace{5mu} + \frac{1}{R^{\text{wz}}} (T^{\text{wu}}_{t} - T^{\text{zu}}_{t}) \Bigr), \quad \forall{t} < J-1 \label{P2:StateSpace1} \\
        T^{\text{zl}}_{t+1} & = T^{\text{zl}}_{t} + dt \cdot \frac{1}{C^{\text{zl}}}\Bigl( \frac{1}{R^{\text{zuzl}}} (T^{\text{zu}}_{t} - T^{\text{zl}}_{t}) \notag                                     \\ & \mspace{5mu} + \frac{1}{R^{\text{wz}}} (T^{\text{wl}}_{t} - T^{\text{zl}}_{t}) \Bigr), \quad \forall{t} < J-1 \label{P2:StateSpace2} \\
        T^{\text{wu}}_{t+1} & = T^{\text{wu}}_{t} + dt \cdot \frac{1}{C^{\text{wu}}}\Bigl( (1-\mathbbm{1}^{\text{lid}}) \frac{1}{R^{\text{wua},\text{off}}} (T^{\text{a}} - T^{\text{wu}}_{t}) \notag + \\ & \mspace{5mu} \mathbbm{1}^{\text{lid}} \frac{1}{R^{\text{wua},\text{on}}} (T^{\text{a}} - T^{\text{wu}}_{t}) + \frac{1}{R^{\text{ww}}} (T^{\text{wl}}_{t} - T^{\text{wu}}_{t}) \notag \\ & \mspace{5mu} + \frac{1}{R^{\text{wz}}} (T^{\text{zu}}_{t} - T^{\text{wu}}_{t}) + p^{\text{u}}_{t} \Bigr), \quad \forall{t} < J-1 \label{P2:StateSpace3} \\
        T^{\text{wl}}_{t+1} & = T^{\text{wl}}_{t} + dt \cdot \frac{1}{C^{\text{wl}}}\Bigl( \frac{1}{R^{\text{wla}}} (T^{\text{a}} - T^{\text{wl}}_{t}) \notag                                           \\ & \mspace{5mu} + \frac{1}{R^{\text{ww}}} (T^{\text{wu}}_{t} - T^{\text{wl}}_{t})  \notag \\
        + & \mspace{5mu} \frac{1}{R^{\text{wz}}} (T^{\text{zl}}_{t} - T^{\text{wl}}_{t}) + p^{\text{l}}_{t} \Bigr), \quad \forall{t} < J-1 \label{P2:StateSpace4} \\
        T^{\text{zu, Base}}_{t+1} & = T^{\text{zu, Base}}_{t} + dt \cdot \frac{1}{C^{\text{zu}}}\Bigl( \frac{1}{R^{\text{zuzl}}} (T^{\text{zl, Base}}_{t} - T^{\text{zu, Base}}_{t}) \notag                                     \\ & \mspace{5mu} + \frac{1}{R^{\text{wz}}} (T^{\text{wu, Base}}_{t} - T^{\text{zu, Base}}_{t}) \Bigr), \quad \forall{t} < J-1 \label{P2:StateSpaceBase1} \\
        T^{\text{zl, Base}}_{t+1} & = T^{\text{zl, Base}}_{t} + dt \cdot \frac{1}{C^{\text{zl}}}\Bigl( \frac{1}{R^{\text{zuzl}}} (T^{\text{zu, Base}}_{t} - T^{\text{zl, Base}}_{t}) \notag                                     \\ & \mspace{5mu} + \frac{1}{R^{\text{wz}}} (T^{\text{wl, Base}}_{t} - T^{\text{zl, Base}}_{t}) \Bigr), \quad \forall{t} < J-1 \label{P2:StateSpaceBase2} \\
        T^{\text{wu, Base}}_{t+1} & = T^{\text{wu, Base}}_{t} + dt \cdot \frac{1}{C^{\text{wu}}}\Bigl( (1-\mathbbm{1}^{\text{lid}}) \frac{1}{R^{\text{wua},\text{off}}} (T^{\text{a}} - T^{\text{wu, Base}}_{t}) \notag  \\ & \mspace{5mu} + \mathbbm{1}^{\text{lid}} \frac{1}{R^{\text{wua},\text{on}}} (T^{\text{a}} - T^{\text{wu, Base}}_{t}) + \frac{1}{R^{\text{ww}}} (T^{\text{wl, Base}}_{t} - T^{\text{wu, Base}}_{t}) \notag \\ & \mspace{5mu} + \frac{1}{R^{\text{wz}}} (T^{\text{zu, Base}}_{t} - T^{\text{wu, Base}}_{t}) + p^{\text{u, Base}}_{t} \Bigr), \quad \forall{t} < J-1 \label{P2:StateSpaceBase3} \\
        T^{\text{wl, Base}}_{t+1} & = T^{\text{wl, Base}}_{t} + dt \cdot \frac{1}{C^{\text{wl}}}\Bigl( \frac{1}{R^{\text{wla}}} (T^{\text{a}} - T^{\text{wl, Base}}_{t}) \notag                                           \\ & \mspace{5mu} + \frac{1}{R^{\text{ww}}} (T^{\text{wu, Base}}_{t} - T^{\text{wl, Base}}_{t}) \notag \\ & \mspace{5mu}  + \frac{1}{R^{\text{wz}}} (T^{\text{zl, Base}}_{t} - T^{\text{wl, Base}}_{t}) + p^{\text{l, Base}}_{t} \Bigr), \quad \forall{t} < J-1 \label{P2:StateSpaceBase4} \\
        \lambda_{h}^{\rm{bid}} & - M  (1 - g_{h}) \leq \lambda_{h}^{\rm{b}} - \lambda_{h}^{\rm{s}} \leq \lambda_{h}^{\rm{bid}} + M  g_{h}, \ \forall{h}                               \label{con_bid:subeq1} \\
        p^{\rm{b}, \uparrow}_{h} & \leq \phi_{h}  \mathbbm{1}^{\lambda^{\rm{b}}_{h} > \lambda^{\rm{s}}_{h}}, \  \forall{h}                          \label{con_bid:subeq3}                                                \\
        p^{\rm{b}, \uparrow}_{h} + s_{h} & \geq \phi_{h}  \mathbbm{1}^{\lambda^{\rm{b}}_{h} > \lambda^{\rm{s}}_{h}},  \ \forall{h}            \label{con_bid:subeq4}                                               \\
         -g_{h}  M & \leq \phi_{h} \leq g_{h}  M,\  \forall{h}                                   \label{con_bid:subeq5}                                                                                                            \\
         & -(1 - g_{h})  M \leq \phi_{h} - p^{\rm{r},\uparrow}_{h} \leq (1 - g_{h}) M,  \forall{h}\                                                                                    \label{con_bid:subeq7} \\
        %
           & \ p_{h}^{\rm{q}} = P^{\rm{q},\rm{Base}}_{h} - p^{\rm{q},\rm{b}, \uparrow}_{h} + p^{\rm{q},\rm{b}, \downarrow}_{h}, \                                                                                                  \forall{h, q}                                                                             \label{power:6q}                                                                                                                                                                                                                           \\
        \  & p^{\rm{q},\rm{b}, \uparrow}_{h} \leq p^{\rm{r}}_h \mathbbm{1}_{h}^{\lambda^{\rm{b}}_{h} > \lambda^{\rm{s}}_{h}} , \                                                                            \forall{h,q}                                                                             \label{power:8q}                                                                                                                                                                                                          \\
        \  & p^{\rm{q},\rm{b}, \uparrow}_{h} \leq u_{h}^{\rm{q},\uparrow} \big(P^{\rm{q},\rm{Base}}_{h} - P^{\rm{q},\rm{Min}}\big) , \                                                                                                       \forall{h,q}                                                                             \label{power:9q}                                                                                                                                                                                                                      \\
        \  & p^{\rm{q},\rm{b}, \downarrow}_{h} \leq u^{\rm{q},\downarrow}_{h} \big(P^{\rm{q},\rm{Nom}} -P^{\rm{q},\rm{Base}}_{h}\big), \                                                                                              \forall{h,q}                                                                             \label{power:10q}                                                                                                                                                                                                                            \\
        \  & P^{\rm{q},\rm{Min}} \leq p_{h}^{\rm{q}} \leq P^{\rm{q},\rm{Nom}}, \                                                                                                                                           \forall{h,q}                                                                             \label{power:11q}                                                                                                                                                                                                                                         \\
        \  & 0 \leq s_{h}^{\rm{q}} \leq P^{\rm{q},\rm{Base}}_{h}, \                                                                                                                                                   \forall{h,q}                                                                             \label{power:12q}                                                                                                                                                                                                                                       \\
        \  & p^{\rm{q},\rm{b}, \downarrow}_{h} \geq 0.10 \  u^{\rm{q},\downarrow}_{h} \big(P^{\rm{q},\rm{Nom}} - P^{\rm{q},\rm{Base}}_{h}\big), \                                                                                  \forall{h,q}                                                                             \label{power:14q}                                                                                                                                                                                                                               \\
        \  & u_{h-1}^{\rm{q},\uparrow} - u_{h}^{\rm{q},\uparrow} + y_{h}^{\rm{q},\uparrow} - z_{h}^{\rm{q},\uparrow} = 0, \    \forall{h>1,q},                                                                                                         \label{aux:1}                                                                                                                                                                                                                                                                                         \\
        \  & y_{h}^{\rm{q},\uparrow} + z_{h}^{\rm{q},\uparrow} \leq 1 \                                                             \forall{h,q}                                                                                                                                                                     \label{aux:2}                                                                                                                                                                                                                           \\
        \  & u_{h-1}^{\rm{q},\downarrow} - u_{h}^{\rm{q},\downarrow} + y_{h}^{\rm{q},\downarrow} - z_{h}^{\rm{q},\downarrow} = 0, \                                                                                                                                                                                                                                                    \forall{h>1,q},                                                                                                                                        \label{aux:3} \\
        \  & y_{h}^{\rm{q},\downarrow} + z_{h}^{\rm{q},\downarrow} \leq 1 \                                                         \forall{h,q}                                                                                                                                                                     \label{aux:4}                                                                                                                                                                                                                           \\
        \  & u_{h}^{\rm{q},\uparrow} + u_{h}^{\rm{q},\downarrow} \leq 1 \                                                           \forall{h,q}                                                                                                                                                                     \label{aux:5}                                                                                                                                                                                                                           \\
        \  & y_{h}^{\rm{q},\uparrow} + y_{h}^{\rm{q},\downarrow} \leq 1 \                                                           \forall{h,q}                                                                                                                                                                                   \label{aux:6}                                                                                                                                                                                                             \\
        \  & z_{h}^{\rm{q},\uparrow} + z_{h}^{\downarrow} \leq 1 \                                                           \forall{h,q} \label{aux:7}                                                                                                                                                                                                                                                                                                                                                                                               \\
        \  & y^{\rm{q},\downarrow}_{h} \geq z^{\rm{q},\uparrow}_{h}, \                                                                                                                                                                                                                                                                 \forall{h,q} \label{rebound:1}                                                                                                                                                                                     \\
        \  & \sum_{t=4(h-1)}^{4 h} T^{\rm{q}}_{t} - T^{\rm{q}, \rm{Base}}_{t} \geq \big( z^{\rm{q},\downarrow}_{h} -1\big)  M,  \  \forall{h>1,q} \label{rebound:3}                                                                                                                                                                                                                                                                                                                                                                                 \\
        \  & \sum_{k=1}^{h} y^{\rm{q},\downarrow}_{k} \leq y^{\rm{q},\uparrow}_{k}, \ \forall{h,q}. \label{up_reg_first}
    \end{align}
\end{subequations}

The objective function (\ref{P2:1}) maximizes the expected flexibility value of the zinc furnace.
%

Constraints \eqref{P2:2} represent the sum of zone power variables, i.e., total power, balancing power, and slack variable. The slack variable $s_h$ represents power not delivered as promised. Constraints \eqref{P2:3}-\eqref{P2:33} represent sum of total balancing power as the sum of balancing power in both zones.

Aligned with (\ref{eq:StateSpaceModel}), constraints \eqref{P2:StateSpace1}-\eqref{P2:StateSpace4} are the state-space model for the zinc and furnace wall temperature dynamics. Similarly, \eqref{P2:StateSpaceBase1}-\eqref{P2:StateSpaceBase4} include the baseline temperatures for the zinc and furnace wall, and model temperature dynamics for the baseline power. 
Recall in case the hour index $h$ runs from 1 to 24, index $t$ runs from 1 to $J=1440$.

Constraint \eqref{con_bid:subeq1}-\eqref{con_bid:subeq7} represent the McCormick relaxation of activation conditions for mFRR, i.e., that the zinc furnace should activate its capacity whenever there is a positive reserve bid, $p^{r}_{h} > 0$, and when the regulation power bid is below the balancing price, $\lambda_{h}^{\rm{bid}} < \lambda_{h}^{\rm{b}}$, given that the balancing price is strictly greater than the spot price, $\lambda_{h}^{\rm{b}} > \lambda_{h}^{\rm{s}}$, i.e., whenever the system required balancing. See more details in \cite{gade2023load}.

Constraint \eqref{power:6q} sets the real-time power consumption $p_{h}^{\rm{q}}$ for both zones equal to the baseline power $P^{\rm{q},\rm{Base}}_{h}$ unless there is up-regulation $p^{\rm{q},\rm{b}, \uparrow}_{h}$ or down-regulation $p^{\rm{q},\rm{b}, \downarrow}_{h}$.
%
%
Constraint (\ref{power:8q}) ensures that up-regulation is zero when there is no need for up-regulation, and at the same time binds it to the reservation power.
Constraint (\ref{power:9q}) includes the binary variable $u^{\rm{q},\uparrow}_{h}$, indicating whether the zinc furnace is up-regulated in hour $h$. This constraint ensures that up-regulation is zero whenever $u^{\rm{q},\uparrow}_{h} = 0$, and otherwise restricted to the maximum up-regulation service $P^{\rm{q},\rm{Base}}_{h}-P^{\rm{q},\rm{Min}}$ that can be provided. Note that $P^{\rm{q},\rm{Min}}$ is the minimum consumption level of the zinc furnace for zone $q$.
Constraint (\ref{power:10q}) works similarly for down-regulation. Note that the binary variable $u^{\rm{q},\downarrow}_{h}$ indicates whether down-regulation happens, whereas $P^{\rm{q}\rm{Nom}}$ is the nominal (maximum) consumption level of the zinc furnace.
Constraint (\ref{power:11q}) restricts the power consumption to lie within the minimum and nominal rates for each zone.
Constraint (\ref{power:12q}) binds the slack variable $s_{h}^{\rm{q}}$, representing the service not delivered as promised.
Constraint (\ref{power:14q}) ensures that down-regulation is equal to at least 10\% of the down-regulation capacity.
%
%
Constraints \eqref{aux:1}-\eqref{aux:7} define auxiliary binary variables $y^{\rm{q},\uparrow}_{h}$, $y^{\rm{q},\downarrow}_{h}$, $z^{\rm{q},\uparrow}_{h}$, and $z^{\rm{q},\downarrow}_{h}$, identifying transitions from/to up-regulation and down-regulation.
During all hours with up-regulation, $y^{\rm{q},\uparrow}_{h}=1$. In the hour that up-regulation is stopped, $z^{\rm{q},\uparrow}_{h}$ is 1. There is a similar definition for $y^{\rm{q},\downarrow}_{h}$ and $z^{\rm{q},\downarrow}_{h}$ related to down-regulation.
%
%
Constraints \eqref{rebound:1}-\eqref{rebound:3} control the rebound behavior such that the rebound finishes when the temperature is below the baseline temperature. Note that $M$ is a sufficiently big positive constant such that the zinc temperature is allowed to deviate from the baseline. Also, they ensure that the rebound happens right after up-regulation.
Lastly, (\ref{up_reg_first}) ensures that up-regulation happens first. This makes sense since it impossible (or at least difficult) to anticipate potential up-regulation events in the power system. As such, it does not make sense to pre-cool (or pre-heat) a TCL in the context of mFRR.

\vfill

\end{document}